\documentclass[twocolumn,showpacs,preprintnumbers,amsmath,amssymb,prl,
]{revtex4}
\usepackage{graphicx}
\usepackage{dcolumn}
\usepackage{bm}
\usepackage{amsmath}
\usepackage{hyperref}
\usepackage[dvips]{color}

\def\rnum#1{\expandafter{%
\romannumeral #1}}
\def\Rnum#1{\uppercase\expandafter{%
\romannumeral #1}}

\newcommand{\bol}[1]{\boldsymbol #1}

\begin{document}

\title{Quasi Long Range Order of Defects in Frustrated 
Antiferromagnetic Ising Models on Spatially Anisotropic Triangular Lattices}

\author{Masahiro Sato, Naoyuki Watanabe, and Nobuo Furukawa}
\affiliation{Department of Physics and Mathematics, Aoyama Gakuin University, 
Fuchinobe 5-10-1, Sagamihara, 229-8558 Japan} %\\

\date{\today}

\begin{abstract}
It is known that there is no phase transition down to zero 
temperature in the antiferromagnetic Ising model on spatially anisotropic 
triangular lattices, in which the exchange coupling of one direction 
is stronger than those of other two directions. In the model, the 
low-temperature physics is governed by domain-wall excitations (defects) 
residing on bonds of the strong-coupling direction. 
In this letter, we show that an additional small attractive interaction 
between defects (a ferromagnetic next-nearest-neighbor interaction 
in the weak-coupling direction) leads to a Berezinskii-Kosterlitz-Thouless 
(BKT) transition at a finite temperature, by performing the Monte Carlo 
simulation. The BKT phase can be viewed as the phase with a quasi 
long-range order of defects. 
We determine the phase diagram in a wide parameter regime and 
argue the phase structure from statistical-mechanics and 
field-theory viewpoints.\\

\noindent
{\bf Keywords}: {\it Defects, Frustration, Ising model, 
Berezinskii-Kosterlitz-Thouless transition, Conformal field theory}
\end{abstract}

%\pacs{75.10.-b, 75.10.Jm, 75.10.Pq, 75.30.Fv, 75.40.Gb}
%75.10.-b General theory and models of magnetic ordering
%75.10.Jm  Quantized spin models, including quantum spin frustration
%75.10.Pq Spin chain models
%75.40.Gb Dynamic properties (dynamic susceptibility, spin waves,
%spin diffusion, dynamic scaling, etc.)
%75.30.Kz Magnetic phase boundaries (including classical and quantum
%magnetic transitions, metamagnetism, etc.)
%75.40.Cx Static properties (order parameter, static susceptibility,
%heat capacities, critical exponents, etc.)
%75.30.Fv Spin-density waves

\maketitle

%%%%%%%%%%%%%%%%%%%%%%%%%%%%%%%%%%%%%%%%%%%%%%%%%%%%%%%
%%%%%%%%%%%%%%%%%%%%%%%%%%%%%%%%%%%%%%%%%%%%%%%%%%%%%%%
%%%%%%%%%%%%%%%%%%%%%%%%%%%%%%%%%%%%%%%%%%%%%%%%%%%%%%%
\textit{Introduction}.$-$
Antiferromagnets on triangular lattices~\cite{Canada} have long been studied 
as representative systems with geometric frustration~\cite{Lacroix,Diep}. 
In frustrated magnets, spins cannot take their lowest-energy configuration 
and hence macroscopic many-body states are almost degenerate 
at sufficiently low temperatures. 
This is a main reason why various frustrated magnets 
often lead to intriguing phenomena. 
For continuous or quantum spin systems on triangular lattices, 
spin chiralities~\cite{Kawamura} 
such as ${\bol S}_{\bol r}\times{\bol S}_{\bol r'}$ 
have offered interesting topics. In addition to chiralities, recently 
multiferroic properties~\cite{multiferro}, 
spin nematic orders~\cite{Lacroix2}, and 
spin-liquid states~\cite{Balents} 
have been vividly explored as novel many-body states or phenomena.

\begin{figure}
\includegraphics[width=8cm]{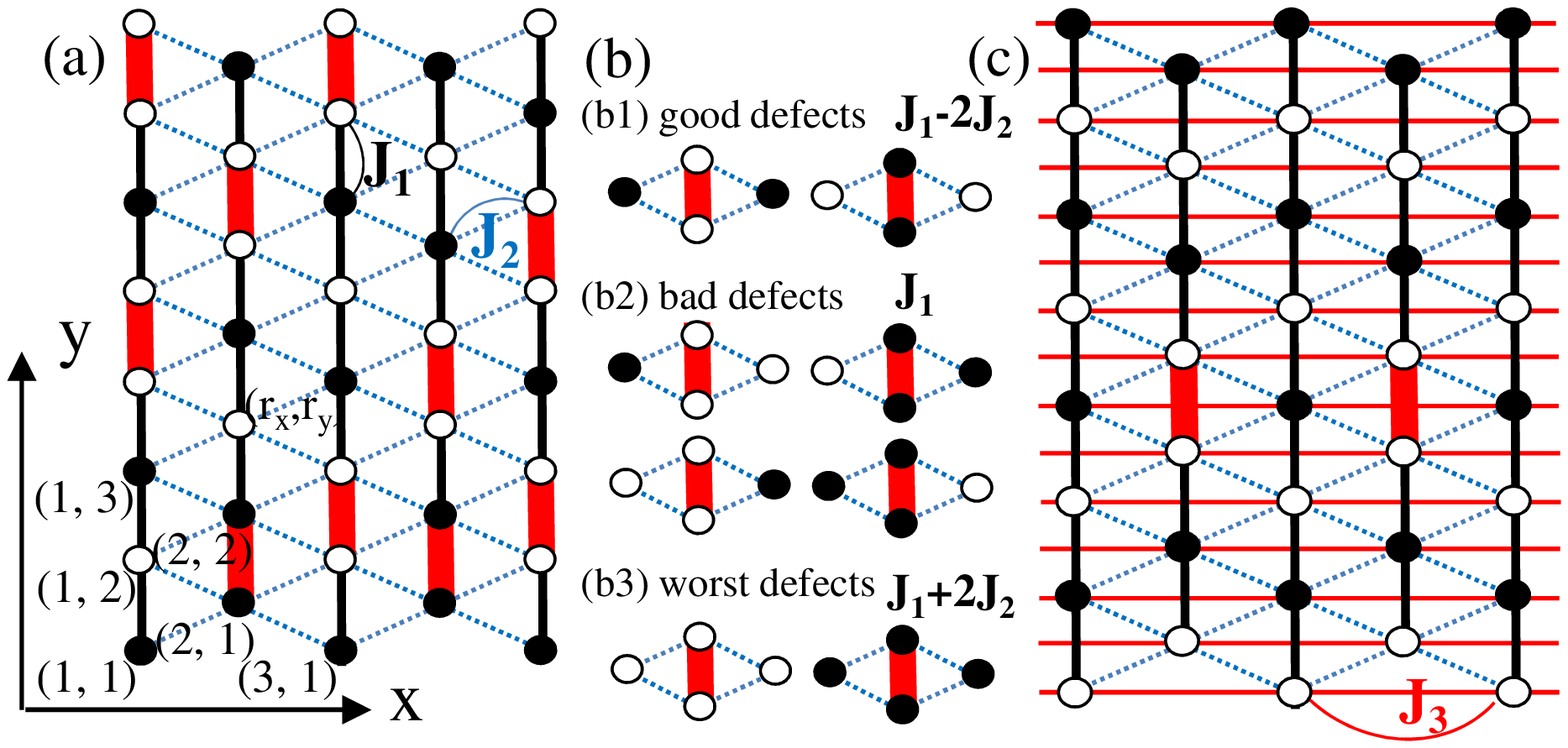}
\caption{(Color online) 
(a) AF Ising model ${\cal H}_{J_1-J_2}$ (\ref{eq:TriangleIsing_J1J2}) 
on a triangular lattice, in which white and black circles respectively denote 
up and down spins ($S_{\bol r}=1$ and -1). 
Red thick bonds show the positions of defects. 
(b) Three types of defects (b1)-(b3) 
in the model (\ref{eq:TriangleIsing_J1J2}). Values $J_1-2J_2$, $J_1$, and 
$J_1+2J_2$ are the energies of defects measured from 
the state without any bond. 
%We note that a single spin flip in N\'eel ordered 
%chain creates two defects. 
(c) Modified Ising model ${\cal H}_{J_1-J_2-J_3}$ 
(\ref{eq:TriangleIsing_J1J2J3}) with an additional next-nearest-neighbor
coupling $J_3$.}
\label{fig:lattice}
\end{figure}

Triangular antiferromagnets made from discrete spin degrees of freedom 
(Ising, clock, Potts models, etc.)~\cite{Diep2} are also interesting 
frustrated systems. They are also useful as simple models of, 
for instance, charge orderings, electric dipoles, lattice gases, 
and crystal growths. 
For these systems, their macroscopically degenerate 
ground states can be often counted out, but small thermal or 
quantum fluctuations in the ground-state manifold can 
induce unexpected phenomena. Hence effects of those fluctuations have 
offered an attractive research field. 
In this paper, we focus on antiferromagnetic (AF) spin-$\frac{1}{2}$ 
Ising systems on triangular lattices. 
%which belong to a typical kind of models 
%for frustrated magnets with discrete degrees of freedom. 
We start with the following Hamiltonian with a spatial anisotropy: 
\begin{eqnarray}
{\cal H}_{J_1-J_2}&=& J_1\sum_{\langle {\bol r},{\bol r'}\rangle_y}
S_{\bol r} S_{\bol r'} 
+J_2\sum_{\langle {\bol r},{\bol r'}\rangle_x}
S_{\bol r}S_{\bol r'}, 
\label{eq:TriangleIsing_J1J2}
\end{eqnarray}
where $S_{\bol r}$ is the Ising spin on site ${\bol r}=(r_x,r_y)$ 
and take values $\pm 1$, and $J_1>0$ and $J_2>0$ respectively 
denote the AF exchange coupling constants along one direction 
of the triangular lattice and along other two directions. 
This model is illustrated in Fig.~\ref{fig:lattice}(a). 
As shown in the figure, 
the symbols $\langle {\bol r},{\bol r'}\rangle_x$ and 
$\langle {\bol r},{\bol r'}\rangle_y$ respectively stand for nearest-neighbor 
Ising spin pair along the $x$ and $y$ directions. The ratio $J_2/J_1$ measures 
the spatial anisotropy. The point $J_2/J_1=0$ corresponds to 
decoupled AF-$J_1$ Ising chains, while $J_2/J_1=1$ is the 
spatially-isotropic triangular Ising model. We will consider the wide range 
$0<J_2/J_1\leq 1$ hereafter. 

The study of the spatially anisotropic Ising 
model~(\ref{eq:TriangleIsing_J1J2}) has a long history~\cite{Diep2}. 
Wannier has studied thermodynamic properties of the isotropic 
model~(\ref{eq:TriangleIsing_J1J2}) with $J_1=J_2$, 
by using transfer matrix method~\cite{Wannier}. 
He has accurately computed the residual entropy for the macroscopically 
degenerate ground states, and shown that the internal energy 
has no singular point as a function of temperature $T$. 
The zero-temperature critical properties~\cite{Stephenson,Blote,Nienhuis,
Blote2,Otsuki} and effects of some perturbations~\cite{Mekata,Landau,Takayama,
Slotte,Fujiki,Kitatani,Qian} have been studied as well. 
For $J_1>J_2>0$, the ground states consist of all of the states 
where Ising chains along the $y$ direction are independently N\'eel ordered. 
In fact, when N\'eel ordered chains align to the $y$ direction, 
the inter-chain energy is canceled out. 
As a result, the ground states are semi-macroscopically 
degenerate~\cite{comment2}. 
Houtappel has examined the general anisotropic case 
with $J_1\neq J_2$, and derived the exact expression 
of the partition function~\cite{Houtappel}. 
Recently, Hotta, {\it et. al.}~\cite{Hotta,Hotta2} 
have studied thermodynamic properties of the 
model~(\ref{eq:TriangleIsing_J1J2}) with $J_2/J_1<1$, 
by using the exact solution of Ref.~\onlinecite{Houtappel} 
and Monte Carlo (MC) simulation. 
They have confirmed that there is no phase transition in the plane 
of $T$ and $J_2/J_1(<1)$. They have pointed out that 
the low-energy physics of the anisotropic 
model~(\ref{eq:TriangleIsing_J1J2}) with $J_2/J_1<1$ are governed by 
domain-wall excitations (defects) as shown in Fig.~\ref{fig:lattice}(a) and 
(b). The defects are the ferromagnetic bonds cutting 
N\'eel ordered state of each Ising chain, and they are classified into 
three kinds in the energetic sense, as in Fig.~\ref{fig:lattice}(b); 
Good, bad, and worst defects. In a sufficiently low-temperature regime, 
bad and worst defects disappear and only good defects survive. 
The inter-chain coupling $J_2$ develops a short-range correlation 
between defects along the $x$ direction, 
but it does not induce any phase transition.

These results make us infer that defects form a ordering pattern 
(crystallization of defects) as in electrons of Wigner crystals 
if we introduce a small interaction between defects 
in the anisotropic model~(\ref{eq:TriangleIsing_J1J2}) with $J_2/J_1<1$. 
Following this expectation, we will study the possibility 
of defect crystallization in this paper. 
To this end, we introduce an additional 
ferromagnetic interaction $J_3<0$ to the $J_1$-$J_2$ 
model~(\ref{eq:TriangleIsing_J1J2}). The new Hamiltonian is given by 
\begin{eqnarray}
{\cal H}_{J_1-J_2-J_3}&=& {\cal H}_{J_1-J_2}
+J_3\sum_{\langle\langle {\bol r},{\bol r'}\rangle\rangle_x}
S_{\bol r}S_{\bol r'}.
\label{eq:TriangleIsing_J1J2J3}
\end{eqnarray}
The $J_3$ term is a next-nearest-neighbor interaction along the $x$ direction 
and is illustrated in Fig.~\ref{fig:lattice}(c). 
The panel (c) shows that a ferromagnetic $J_3$ term plays a role of an 
attraction between two neighboring good defects.

%%%%%%%%%%%%%%%%%%%%%%%%%%%%%%%%%%%%%%%%%%%%%%%%%%%%%%%
%%%%%%%%%%%%%%%%%%%%%%%%%%%%%%%%%%%%%%%%%%%%%%%%%%%%%%%
%%%%%%%%%%%%%%%%%%%%%%%%%%%%%%%%%%%%%%%%%%%%%%%%%%%%%%%
\textit{Numerical analysis}.$-$ 
We will apply the standard heat-bath MC method to 
the $J_1$-$J_2$-$J_3$ model~(\ref{eq:TriangleIsing_J1J2J3}). 
Both lengths of the $x$ and $y$ directions, $L_x$ and $L_y$, are set to 
$L$, and the periodic boundary condition for both directions 
are imposed. The system size $L$ is increased up to 128. 
Taking the thermal average, we use $O(10^7)$ MC samplings. 
Hereafter, we will set $J_3/J_1=-0.1$ in the 
model~(\ref{eq:TriangleIsing_J1J2J3}) in order to 
see effects of a "small" additional interaction $J_3$ on defects of the 
$J_1$-$J_2$ model~(\ref{eq:TriangleIsing_J1J2}).

\begin{figure}
\includegraphics[width=8cm]{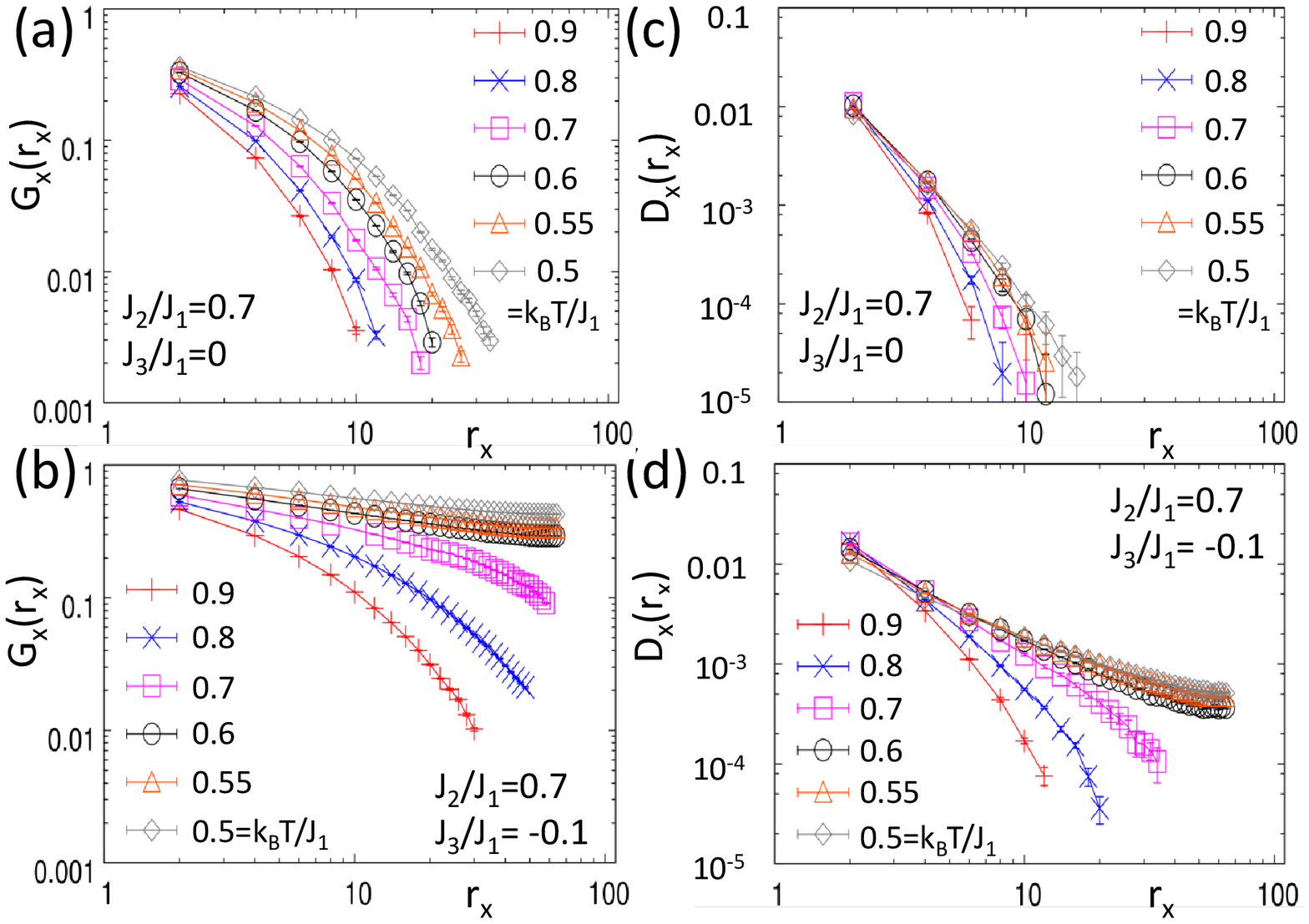}
\caption{(Color online) 
(a)(b) Spin correlation functions $G_{x}(r_x)$ 
of the models~(\ref{eq:TriangleIsing_J1J2}) and 
(\ref{eq:TriangleIsing_J1J2J3}). 
(c)(d) Defect correlation functions $D_{x}(r_x)$ of the 
models~(\ref{eq:TriangleIsing_J1J2}) and (\ref{eq:TriangleIsing_J1J2J3}). 
We have set $L=128$ and plotted only the results of even $r_x$. 
The amplitudes of odd $r_x$ are very small.}
\label{fig:Correaltion}
\end{figure}

Let us first see temperature $k_BT$ and anisotropy $J_2/J_1$ dependences 
of spin and defect correlation functions along the $x$ direction. 
These two correlators are defined as 
\begin{eqnarray}
G_x(r_x) &=& \langle S_{(1,r_y)}S_{(r_x+1,r_y)}\rangle
-\langle S_{(1,r_y)}\rangle\langle S_{(r_x+1,r_y)}\rangle,\\
D_x(r_x) &=& \langle d_{(1,r_y)}d_{(r_x+1,r_y)}\rangle
-\langle d_{(1,r_y)}\rangle\langle d_{(r_x+1,r_y)}\rangle.
\label{eq:correaltion_x}
\end{eqnarray}
The quantity $d_{\bol r}=(1+S_{\bol r}S_{(r_x,r_y+1)})/2$ stands for 
the defect operator on the bond between $(r_x,r_y)$ and $(r_x,r_y+1)$. 
It takes unity when a defect exists on the bond, 
while it becomes zero otherwise. 
Figure~\ref{fig:Correaltion} shows the $T$ dependence of both 
spin and defect correlation functions in the 
models~(\ref{eq:TriangleIsing_J1J2}) and (\ref{eq:TriangleIsing_J1J2J3}) 
at $J_2/J_1=0.7$. The panels (a) and (c) clearly indicates that 
spin and defect correlation functions decay exponentially 
irrespective of $k_BT$ in the 
model~(\ref{eq:TriangleIsing_J1J2}) without $J_3$ term. This is consistent 
with the fact that the $J_1$-$J_2$ model has no phase transition. 
On the other hand, panels (b) and (d) show that 
the decay fashion of spin and defect correlators changes into an algebraic 
type from an exponential form when $T$ is sufficiently decreased in 
the $J_1$-$J_2$-$J_3$ model. This power-law behavior strongly suggests 
the emergence of a quasi long-range order of defects and spins, 
namely, a Berezinskii-Kosterlitz-Thouless (BKT) 
phase~\cite{Berezinskii,KT,Kosterlitz,Kogut,Kadanoff}. 

\begin{figure}
\includegraphics[width=8cm]{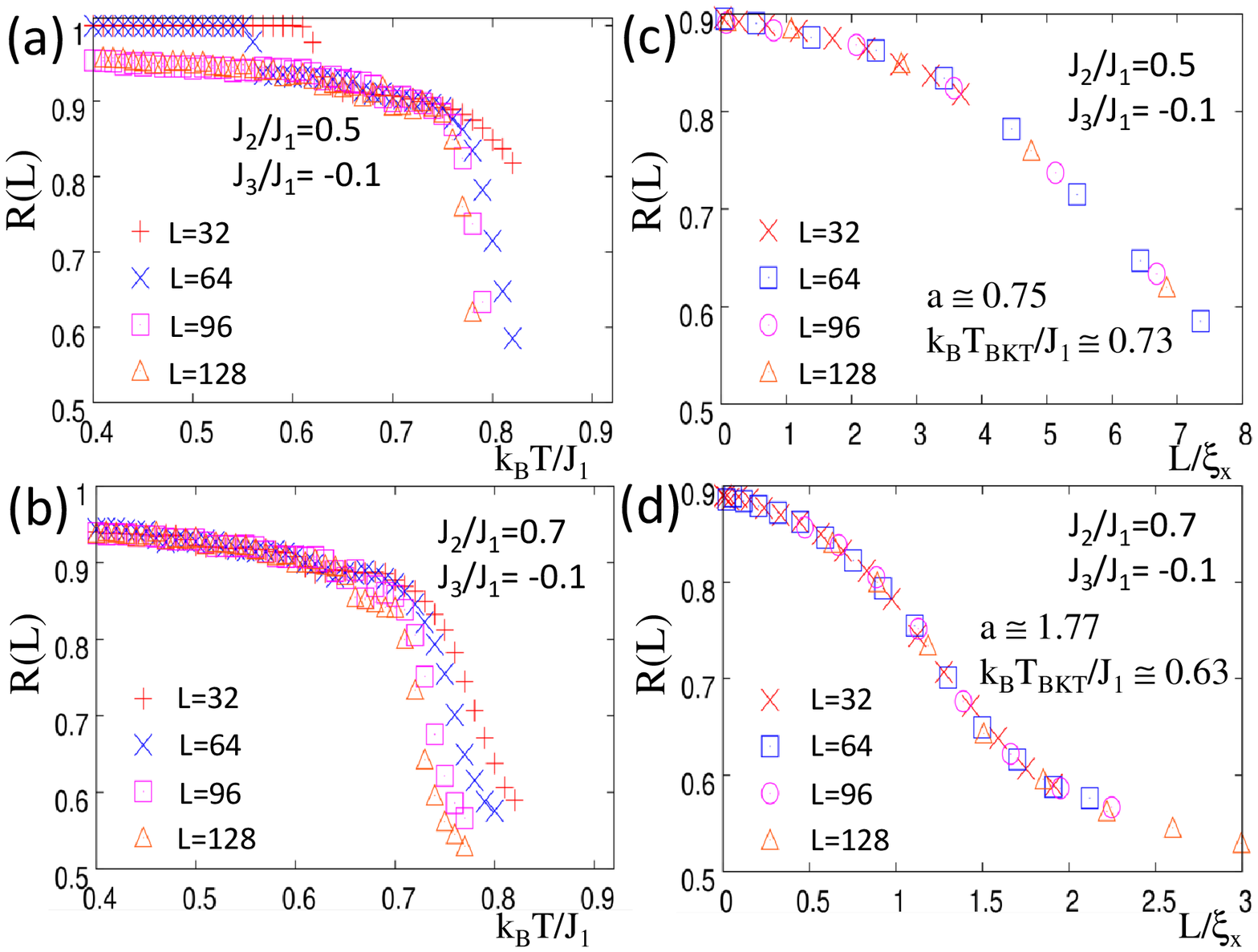}
\caption{(Color online) 
(a)(b) Ratio of two spin correlators $R(L)$ as a function of $T$. 
(c)(d) $R(L)$ as a function of $L/\xi_x$ with 
$\xi_x=e^{a/\sqrt{t}}$. Parameters $a$ and $t$ (i.e., $T_{BKT}$) 
are finely tuned.}
\label{fig:Ratio_Correlators}
\end{figure}

\begin{figure}
\includegraphics[width=8cm]{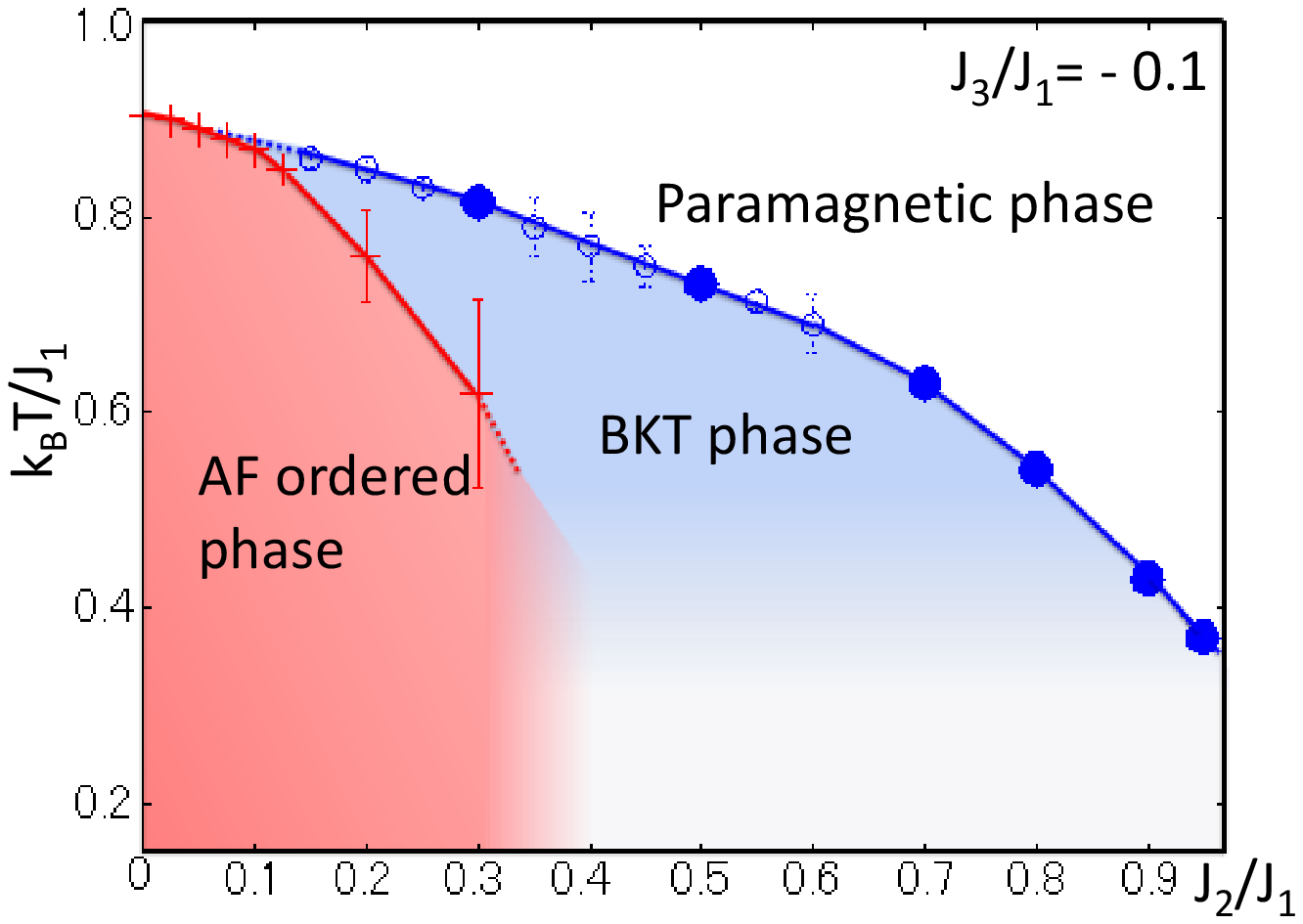}
\caption{(Color online) 
Phase diagram of the model ${\cal H}_{J_1-J_2-J_3}$ 
in the space $(k_BT,J_2)$. Incommensurate ordered phases (see the text) 
might appear between BKT and AF ordered phases.}
\label{fig:Phases}
\end{figure}

To judge whether or not a BKT phase exists and 
to correctly determine the BKT transition point from the MC results, 
we should more carefully treat them 
rather than those of usual second-order transitions. Here, we utilize 
the finite-size scaling based on ratios of two spin correlators, 
that has been developed in Ref.~\onlinecite{Tomita}. 
Around a continuous transition or in a BKT phase of the 
models~(\ref{eq:TriangleIsing_J1J2}) and (\ref{eq:TriangleIsing_J1J2J3}) 
with size $L$ (if they exist), 
a ratio of two spin correlation functions is expected to 
satisfy the following scaling relation:
\begin{eqnarray}
\frac{G_x(r_x,T,L)}{G_x(r_x',T,L)} &=& f(L/\xi_x),
\label{eq:scaling1}
\end{eqnarray}
where $f(L/\xi_x)$ is a function, 
$\xi_x$ is the correlation length of $G_x$, 
and the ratio of two lengths $r_x/r_x'$ is fixed regardless of $T$ and $L$. 
Here, we have explicitly denoted $T$ and $L$ dependences of $G_x$. 
Since $\xi_x$ diverges in a BKT phase, 
the left-hand side of Eq.~(\ref{eq:scaling1}) becomes independent of 
the size $L$ there. In Fig.~\ref{fig:Ratio_Correlators}(a) and (b), 
we plot 
\begin{eqnarray}
R(L)&=& \frac{G_x(L/2,T,L)}{G_x(L/4,T,L)}
\label{eq:scaling2}
\end{eqnarray}
as a function of $T$ in the $J_1$-$J_2$-$J_3$ models 
with different sizes. These two panels uncover that $R(L)$ is almost 
invariant under changing the size $L$ in a finite temperature range. 
This strongly suggests that a BKT phase emerges due to the $J_3$ term. 
In the paramagnetic phase near the BKT transition point, 
$\xi_x$ is expected to behave as 
\begin{eqnarray}
\xi_x &\propto& \exp(a/\sqrt{t})
\label{eq:scaling3}
\end{eqnarray}
where $a$ is a nonuniversal constant, $t=(T-T_{BKT})/T_{BKT}$, and 
$T_{BKT}$ is the BKT transition temperature. 
Therefore, if we plot $R(L)$ as a function of $Le^{-a/\sqrt{t}}$ 
finely tuning parameters $a$ and $T_{BKT}$, 
$R(L)$ with different $L$ and $T$ all align on a single curve 
in the regime $t\ll 1$. Such curves are depicted in 
Fig.~\ref{fig:Ratio_Correlators}(c) and (d). 
Using this scaling method, we determine $T_{BKT}$ 
in the wide range of $0<J_2/J_1<1$ with fixing $J_3/J_1=-0.1$. 
These points $T_{BKT}$ are given by filled circles in 
Fig.~\ref{fig:Phases}.

%%%%%%%%%  phase diagram   %%%%%%%%%%%%%%%%%%%%%%%%

\begin{figure}
\includegraphics[width=8cm]{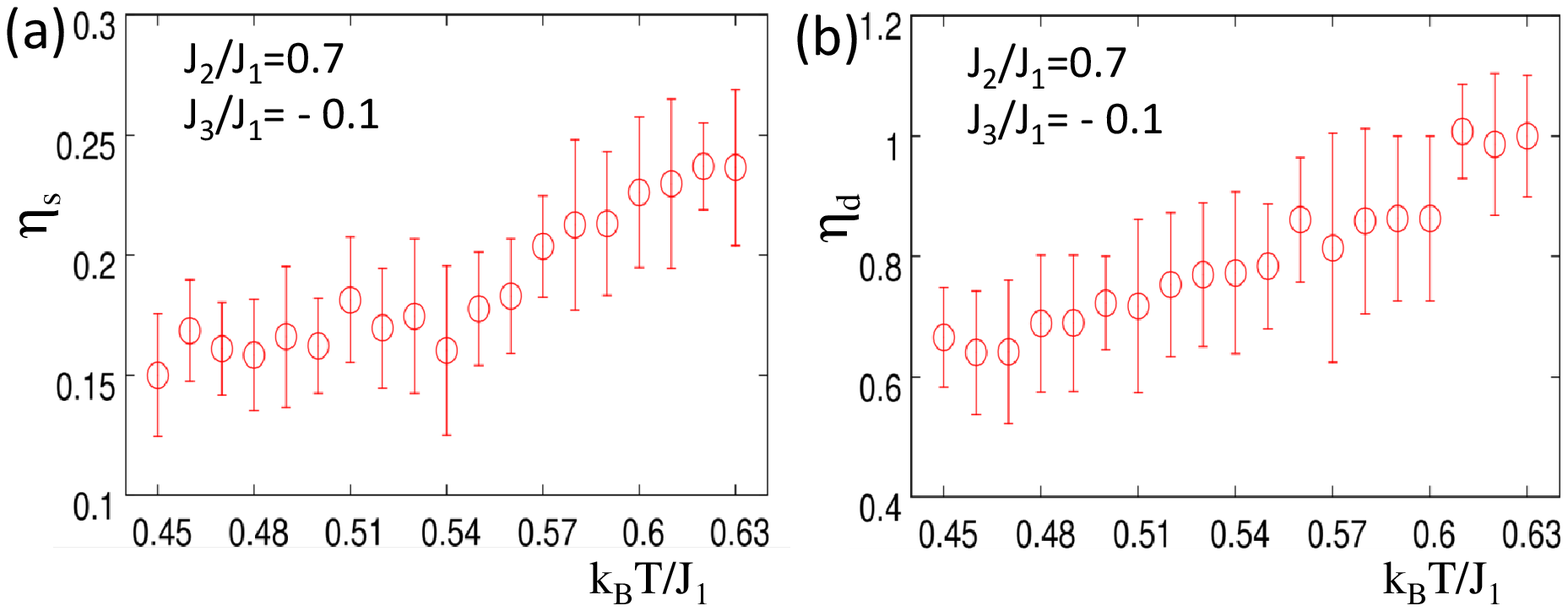}
\caption{(Color online) 
Critical exponents of spin and defect correlation functions 
(a) $\eta_{\rm s}$ and (b) $\eta_{\rm d}$ at $J_2/J_1=0.7$, 
where $k_B T_{BKT}/J_1\approx 0.63$. 
In the calculation of $\eta_{s,d}$, we have used 
the data $G_x(r_x)$ and $D_x(r_x)$ in the range $10<r_x<50$ 
for the $L=128$ systems.}
\label{fig:CriticalExp}
\end{figure}

From spin and defect correlation functions in Fig.~\ref{fig:Correaltion}, 
we can also evaluate their critical exponents $\eta_s$ and $\eta_d$, 
that are defined as
\begin{eqnarray}
G_x(r_x)\sim  r_x^{-\eta_s},&&
D_x(r_x) \sim  r_x^{-\eta_d},
\label{eq:CriticalExp}
\end{eqnarray}
where spatially oscillating factors of $G_x$ and $D_x$ are neglected 
(We again note that all the plotted positions $r_x$ are even 
in Fig.~\ref{fig:Correaltion}). 
The evaluated exponents are plotted in Fig.~\ref{fig:CriticalExp}.
Although there exist large errors in the data of 
Fig.~\ref{fig:CriticalExp}, the result indicates that $\eta_s$ ($\eta_d$) 
takes the value $\sim 1/4$ (unity) at $T=T_{BKT}$, and 
it monotonically decreases with lowering $T$ in the BKT phase. 
We remark that for the well-studied XY model 
on a square lattice, the critical exponent of the spin correlation 
also takes 1/4 at the BKT transition point~\cite{KT,Kosterlitz,Kadanoff}. 
Namely, Ising spins $S_{\bol r}$ of the present $J_1$-$J_2$-$J_3$ model 
exhibit the same asymptotic behavior as XY spins of the XY model 
in the BKT phase. 
From this property of $\eta_{s,d}$, we can also determine $T_{BKT}$ as 
the point where $\eta_s=1/4$ ($\eta_d=1$). 
Temperatures $T_{BKT}$ evaluated from $\eta_{s,d}$ are plotted 
by open circles in Fig.~\ref{fig:Phases}. 
From these results, we conclude that a small $J_3$ term causes a BKT phase, 
and it can be understood as the quasi long-range ordered phase 
of defects and spins.

\begin{figure}
\includegraphics[width=8cm]{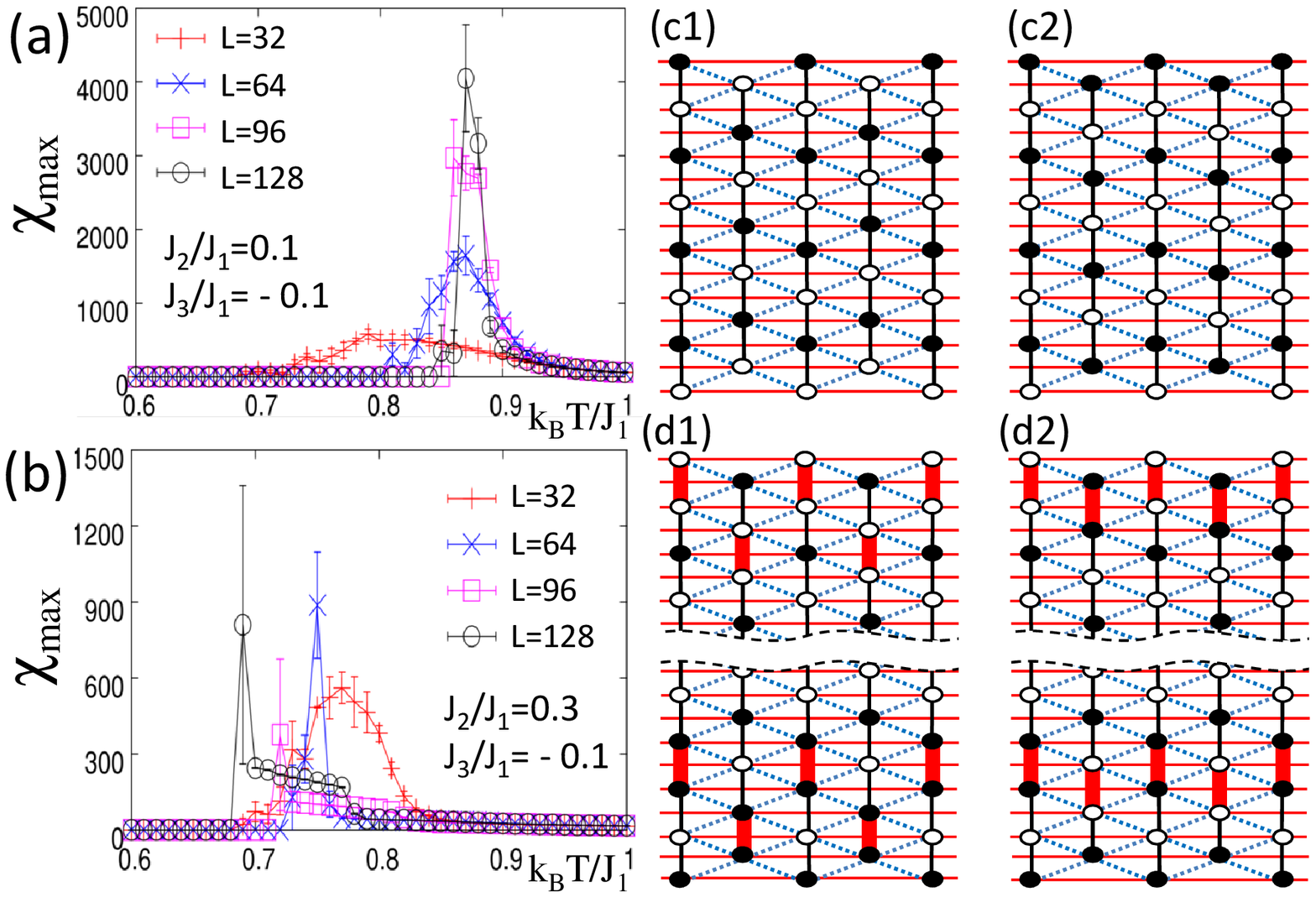}
\caption{(Color online) 
(a)(b) Susceptibility $\chi_{\rm max}$ for the order parameter of 
the AF phase. (c1)(c2) AF ordering patterns. (d1)(d2) Examples of 
possible incommensurate ordering patterns.}
\label{fig:Susceptibility}
\end{figure}

%%%%%%%%%%%%%%%%%%%%%%  AF order    %%%%%%%%%%%%%%%%%%%%%%%%%
%%%%%%%%%%%%%%%%%%%%%%%%%%%%%%%%%%%%%%%%%%%%%%%%%%%%%%%%%%%%%
%%%%%%%%%%%%%%%%%%%%%%%%%%%%%%%%%%%%%%%%%%%%%%%%%%%%%%%%%%%%%
In addition to the BKT phase, we also find an AF ordered phase 
in the $J_1$-$J_2$-$J_3$ model. In fact, at an extreme case with $J_2=0$, 
the system is reduced to two decoupled spatially anisotropic Ising models 
on square lattice that are exactly solvable and have a 
second-order phase transition to a N\'eel phase~\cite{McCoy}. 
At $J_2=0$, the transition temperature $T_c$ is exactly calculated as  
\begin{eqnarray}
\sinh\left(\frac{2J_1}{k_BT_c}\right)
\sinh\left(\frac{2|J_3|}{k_BT_c}\right) &=& 1,
\label{eq:exact_J1J3}
\end{eqnarray}
which leads to $k_BT_c\approx 0.906 J_1$ at $J_3/J_1=-0.1$. 
The ordering pattern is illustrated in Fig.~\ref{fig:Susceptibility}
(c1) and (c2). Hence, their order parameters are respectively defined as 
$m_{1} = L^{-2}\sum_{r_x,r_y}(-1)^{r_y}S_{\bol r}$ and 
$m_{2} = L^{-2}\sum_{r_x,r_y}(-1)^{r_y+r_x+1}S_{\bol r}$. 
%\begin{eqnarray}
%m_{1} = L^{-2}\sum_{r_x,r_y}(-1)^{r_y}S_{\bol r}, \,\,\,\,
%m_{2} = L^{-2}\sum_{r_x,r_y}(-1)^{r_y+r_x+1}S_{\bol r}.
%\label{eq:AForder}
%\end{eqnarray}
The corresponding susceptibilities are given by 
\begin{eqnarray}
\chi_{1,2} &=& \frac{1}{k_BT}
\left(\langle m_{1,2}^2\rangle-\langle m_{1,2}\rangle^2\right). 
\label{eq:AFSusceptibility}
\end{eqnarray}
In Fig.~\ref{fig:Susceptibility}(a) and (b), we draw the MC evaluated 
susceptibility $\chi_{\rm max}={\rm Max}(\chi_1,\chi_2)$ 
as a function of $T$. It exhibits a peak at a certain temperature, 
and the peak becomes sharper with increasing the system size $L$. 
We have confirmed that below the temperature showing the peak, 
an AF order pattern (c1) or (c2) in Fig.~\ref{fig:Susceptibility} 
appears in the MC snap shots. Since the universality class of this 
transition is never known, we determine the transition temperature $T_c$ 
through a naive finite-size scaling for the peak of $\chi_{\rm max}$. 
The evaluated $T_c$ are depicted in Fig.~\ref{fig:Phases}, and one sees 
from it that $T_c$ is lower than $T_{BKT}$ and 
it monotonically decreases with increasing $J_2/J_1$. When $J_2/J_1$ 
exceeds $\sim 0.4$, the evaluated $T_c$ becomes extremely small and 
the error of the finite-size scaling becomes much larger. 
Therefore, we cannot determine $T_c$ in the range $J_2/J_1>0.4$. 
We also see that the peak positions of $\chi_{\rm max}$ seem to 
irregularly change as a function of $L$. 
This strange behavior of $\chi_{\rm max}$ and 
large errors of the finite-size scaling for $T_c$ would be related with 
following things: (i) the MC simulation becomes less reliable 
in a low-temperature regime due to effects of frustration and 
defects, %~\cite{comment1} 
and (ii) the BKT phase above $T_c$ might largely affect the AF ordered 
phase when we consider finite-size systems. 
We will again discuss the nature of the AF transition at $T_c$ later. 
We note that it is difficult to determine whether or not the BKT 
transition curve starts from the $J_2/J_1=0$ axis and 
it coincides with $T_c$ at the axis in Fig.~\ref{fig:Phases}.

\begin{figure}
\includegraphics[width=8cm]{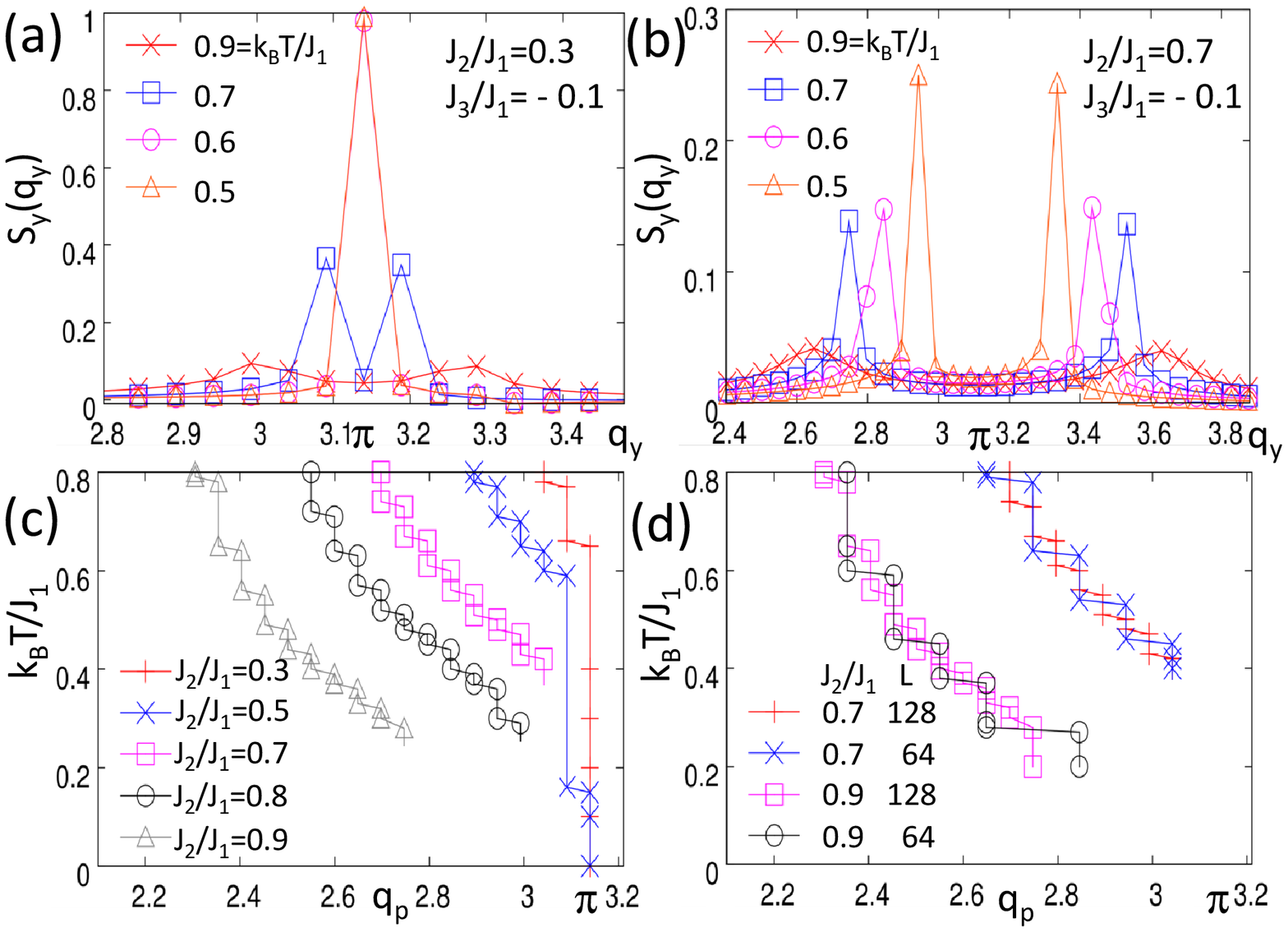}
\caption{(Color online) 
(a)(b) Spin structure factors $S_y(q_y)$. 
(c)(d) Temperature dependence of a peak position $q_p(<\pi)$ of 
$S_y(q_y)$.}
\label{fig:StFactor}
\end{figure}

%%%%%%%%%%%%%%%%%%%%%%%%%%%%%%%%%%%%%%%%%%%%%%%%%%%%%%%%%%%%%
%%%%%%%%%%%%%%%%%%%%%%%%%%%%%%%%%%%%%%%%%%%%%%%%%%%%%%%%%%%%%
%%%%%%%%%%%%%%%%%%%%%%%%%%%%%%%%%%%%%%%%%%%%%%%%%%%%%%%%%%%%%
Now, we turn to correlations along the $y$ direction, 
especially, focusing on the BKT phase. 
Similarly to Eq.~(\ref{eq:correaltion_x}), we define the spin correlation 
function along the $y$ direction $G_y(r_y)=
\langle S_{(r_x,1)}S_{(r_x,r_y+1)}\rangle
-\langle S_{(r_x,1)}\rangle\langle S_{(r_x,r_y+1)}\rangle$. 
Figure~\ref{fig:StFactor}(a) and (b) show the spin structure factor for 
the $y$ direction, that is defined as 
\begin{eqnarray}
S_y (q_y) &=& \frac{1}{L}\sum_{r_y=0}^{L-1}\cos(q_y r_y) G_y(r_y). 
\label{eq:stfactor}
\end{eqnarray}
It has a double peak structure around wave number $q_y=\pi$. 
The temperature dependence of one peak position $q_y=q_p<\pi$ 
are depicted in Fig.~\ref{fig:StFactor}(c) and (d). They show that 
$q_p$ monotonically approaches to $\pi$ with lowering $T$ 
in the whole region $0<J_2/J_1<1$. This is naturally understood 
because $q_p=\pi$ in the AF ordered phase ($T<T_c$), 
and $q_p$ is expected to deviate in proportion to 
the defect density $\langle d_{\bol r}\rangle$ 
that monotonically grows with increasing $T$. 
From Fig.~\ref{fig:StFactor}, we see that $G_y$ is incommensurate 
in the paramagnetic and BKT phases. We have confirmed that 
the "commensurate" spin correlator $G_y(r_y)/\cos(q_pr_y)$ 
exhibits a power-law behavior $\sim r_y^{-\eta_s}$ in the BKT phase. 
This is an additional evidence for the existence of the BKT phase. 
The incommensurate nature of $G_y$ also suggests the 
possibility of single or multiple intermediate phases with an 
incommensurate order between the BKT and AF ordered phases. 
Typical order patterns of possible intermediate phases are given in 
Fig.~\ref{fig:Susceptibility}(d1) and (d2), and they are nothing but 
defect crystals. It is however hard to judge whether or not such phases exist 
since the MC simulation does not work well in $k_BT\ll J_1$ and 
anyone never knows the universality of the phase 
transition to such intermediate phases. 

%%%%%%%  statistical physics and field theory %%%%%%%%%%%%%%%%
%%%%%%%%%%%%%%%%%%%%%%%%%%%%%%%%%%%%%%%%%%%%%%%%%%%%%%%%%%%%%%
%%%%%%%%%%%%%%%%%%%%%%%%%%%%%%%%%%%%%%%%%%%%%%%%%%%%%%%%%%%%%%
\textit{Viewpoints of statistical physics and field theory}.$-$ 
Finally, we discuss the phases and transitions of the 
$J_1$-$J_2$-$J_3$ model from the viewpoints of field theory and 
statistical physics. We have uncovered that the 
$J_1$-$J_2$-$J_3$ model with $J_3/J_1=-0.1$ has two successive 
transitions ($T_{BKT}$ and $T_c$) with lowering $T$ 
at least in the range $0.1<J_2/J_1<0.4$. In addition to the present model, 
a few two-dimensional (2D) classical systems with a BKT phase 
in an intermediate-temperature range have been known. 
$N$-state clock models with $N\geq 5$ are typical 
example~\cite{JKKN,Lecheminant,Kumano}. 
Usually, in such systems, both the transition between paramagnetic and 
BKT phases and that between BKT and ordered phases 
belong to the BKT universality class. 
Order parameters usually grow very slowly in the BKT class.
However, the sharp susceptibility of Fig.~\ref{fig:Susceptibility} 
implies that the transition at $T_c$ is a first- or second-order type. 
To resolve this discrepancy, we can expect some scenarios. 
The first scenario is that the behavior of $\chi_{\rm max}$ 
reduces to a BKT type in the thermodynamic limit. 
The second is (as we mentioned) the possibility of additional 
incommensurate phases between BKT and AF phases. 
If the intermediate-temperature transition is a BKT type, 
the lowest-temperature one may be first- or second-order. 
The third is that the transition between the BKT and AF phases 
is a commensurate-incommensurate (C-IC) type~\cite{Pokrovsky,Giamarchi,Allen}. 
There would also be the possibility that the transition is 
a first-order type. In general, first-order transitions do not 
need to obey a usual picture of the renormalization group. 
The fourth is that the AF order parameter $m_{1,2}$ 
and the defects are effectively decoupled in the long-distance physics. 
To determine the true scenario is an interesting future problem. 
To find the vortex picture of our BKT phase is another important issue. 
In fact, for the BKT (floating) phase of the 2D anisotropic 
next-nearest-neighbor Ising (ANNNI) model, the vortex has been 
defined~\cite{Allen,Villain,Sato,Rastelli}. For the BKT phase of 
the triangular Ising model with isotropic nearest-neighbor and 
next-nearest-neighbor couplings~\cite{Mekata,Takayama,Landau}, 
a vortex picture has also been proposed~\cite{Fujiki}.

BKT phases should be generally described by 
a conformal field theory (CFT) with central charge 
$c=1$~\cite{Yellow,Giamarchi}. 
The effective action for our BKT phase would be given by 
\begin{eqnarray}
{\cal A} = \int dxdy \,\,\,\frac{1}{2}(\partial_\mu\phi)^2 
+g_1 \cos(\sqrt{4\pi K}\phi)  %+g_2 \cos(\sqrt{4\pi}\phi)
+\cdots,
\label{eq:action}
\end{eqnarray}
where $(x,y)$ is the continuous coordinate stemming from the discrete site 
$(r_x,r_y)$, $\phi$ is a scalar field, and $K$ 
is the so-called Tomonaga-Luttinger (TL) liquid parameter~\cite{Giamarchi}. 
In the BKT phase the $g_1$ term with scaling dimension $K$ and 
other perturbations are all irrelevant, and 
the effective action is given by the Gaussian part 
$(\partial_\mu\phi)^2$, while the $g_1$ term becomes relevant ($K<2$) in 
$T>T_{BKT}$ and any quasi long-range correlation disappears. 
As we mentioned, Fig.~\ref{fig:CriticalExp} indicates that Ising spins 
play the same role as spins of 2D XY model in the BKT phase. 
The figure also suggests the relation $\eta_d\approx 4\eta_s$. 
From these two results, we can expect the following correspondence 
between operators of the model~(\ref{eq:TriangleIsing_J1J2J3}) and 
the $c=1$ CFT: 
\begin{eqnarray}
S_{\bol r}\sim \cos({\sqrt{\pi/K}\theta}), && 
d_{\bol r}\sim \cos({2\sqrt{\pi/K}\theta})
\label{eq:VertexOperators}
\end{eqnarray}
where we have neglected spatially oscillating factors. 
%The scaling dimension 
%$\exp(in\sqrt{\pi/K}\theta)$ is $n^2/(4K)$, and 
The field $\theta$ is the dual~\cite{Yellow,Giamarchi} of $\phi$, 
and $\langle e^{in\sqrt{\pi/K}\theta(r)}e^{-in\sqrt{\pi/K}\theta(0)}\rangle 
\sim r^{-n^2/(2K)}$ in the BKT phase. 
Since the low-temperature nature around $T=T_c$ 
has not been revealed enough, it is hard to develop 
the effective theory describing the physics around $T_c$. 
An approach from the $J_2/J_1=0$ limit %(two square-lattice Ising models) 
might be effective to develop such a field theory~\cite{Allen}. 
If the transition at $T_c$ is a C-IC type, a term 
$\partial_y\phi$~\cite{Giamarchi} might be necessary in Eq.~(\ref{eq:action}).

%%%%%%%%%%%%%%%%%%%%%%%%%%%%%%%%%%%%%%%%%%%%%%%%%%%%%%%%%%%%%%%
%%%%%%%%%%%%%%%%%%%%%%%%%%%%%%%%%%%%%%%%%%%%%%%%%%%%%%%%%%%%%%%
%%%%%%%%%%%%%%%%%%%%%%%%%%%%%%%%%%%%%%%%%%%%%%%%%%%%%%%%%%%%%%%
\textit{Conclusion}.$-$ 
In conclusion, we have studied the spatially anisotropic triangular Ising 
model~(\ref{eq:TriangleIsing_J1J2J3}) with an additional $J_3$ term. 
We have shown that a small $J_3<0$ leads to a BKT phase in a wide parameter 
range and its effective field theory has been proposed. The BKT phase 
can be regarded as a quasi long-range ordered phase of defects and spins, 
and an incommensurate spin correlation along the $y$ direction appears there. 
To develop the vortex picture and to reveal 
the low-temperature phases are important remaining issues.

%%%%%%%%%%%%%%%%%%%%%%%%%%%%%%%%%%%%%%%%%%%%%%%%%%%%
%\section{acknowledgment}
%\acknowledgment
\textit{Acknowledgment}.$-$ 
We thank Yuta Kumano and Norikazu Todoroki for valuable discussions. 
This work is supported by KAKENHI (Grant No. 25287088).


\begin{thebibliography}{99}
\bibitem{comment2}
For $J_1>J_2>0$, the ground-state residual entropy is proportional to 
$\sqrt{N}$, where $N$ is the system size.
\end{thebibliography}
\end{document}